\documentclass[twocolumn,showpacs,preprintnumbers,amsmath,amssymb]{revtex4-1}

\usepackage{amssymb}
\usepackage{amsmath}
\usepackage{graphicx}
\usepackage{color}
\usepackage{hyperref}

\newcommand{\be}{\begin{equation}}
\newcommand{\ee}{\end{equation}}
\newcommand{\bea}{\begin{eqnarray}}
\newcommand{\eea}{\end{eqnarray}}
\newcommand{\beal}{\begin{aligned}}
\newcommand{\eeal}{\end{aligned}}

\usepackage{color}

\begin{document}



\title{Minimal surfaces in AdS C-metric}


\author{Hao Xu}
\email{haoxu@mail.nankai.edu.cn}

\affiliation{School of Physics, Nankai University, Tianjin 300071, China}


\date{\today}

\begin{abstract}
We shed some light on the field theory interpretation of C-metric by investigating the minimal surfaces which are homologous to the given boundary regions. The accelerating black holes change the asymptotic structure of the space-time. We focus on the geometry features of the minimal surface and study how deep it reaches into the bulk. The regularized area of the minimal surface is not well defined and we introduce a new quantity $D(m,\theta_0)$, defined as the minimal surface divided by the area of the given boundary region, to study the system.
\end{abstract}


\maketitle

AdS/CFT, or gauge/gravity correspondence, which was discovered by Maldacena in $1990$s, relates the strongly coupled field theory to the classical dynamics of gravity \cite{Maldacena:1997re,Witten:1998qj}. It asserts that the non-gravitational QFT in $d$-dimensions are equivalently described by gravitational interactions in one higher dimensions. AdS/CFT correspondence provides us key insights into various systems and has become a bridge connecting gravity and quantum theory.

The prototype example of the AdS/CFT correspondence is that the type IIB string theory on the product space $AdS_{5}\times S^{5}$ is equivalent to $N=4$ supersymmetric Yang-Mills(SYM) theory on the 4-dimensional boundary. Generally, we may also have more examples on $d$-dimensional CFTs which are dual to the classical gravity models, by viewing the classical gravity as a low energy effective field theory. Matter, which depends on the details of the dual field, can also be allowed. The bulk space-time $M_{d+1}$ satisfies the Einstein's equation, and the CFT$_d$ lives on its boundary $B_d=\partial M_{d+1}$. For the vacuums CFT$_d$, it is dual to the vacuum AdS$_{d+1}$ space-time. For the excited CFT$_d$, it maps to the non-trivial asymptotically AdS space-time determined by Einstein's equation. The simplest case is that the thermal CFT$_d$ is dual to a Schwarzschild-AdS$_{d+1}$ black hole \cite{Witten:1998zw}.

As we have a dictionary between the field theory and asymptotically AdS space-time,
we can try to solve questions in field theory from gravity side, and vice versa.
Giving a holographic CFT$_d$ on the boundary $B_d$, for a spatial region $A$ at some
particular instant in time, we can find a surface $\Sigma$, which is codimensional-2 minimal in the bulk and share the same boundary $\partial A$ with $A$. The minimal
surface satisfies a homology constraint, which demands that $\Sigma$ is smoothly retractable
to the boundary region $A$. In 2006, Ryu and Takayanagi proposed that the quantum entanglement entropy can be directly obtained from minimal surface \cite{Ryu:2006bv,Ryu:2006ef}. The holography entanglement entropy(HEE) is given by the area of the minimal surface in a manner similar to the black hole entropy: $S_{\textsc{A}}=\frac{Area(\Sigma)}{4G}$, where $G$ is the gravitational constant of the bulk theory. This minimal surface also plays an important role in "subregion-subregion" duality. See, e.g., \cite{Nishioka:2009un,Rangamani:2016dms,Hung:2011xb,Dong:2013qoa,
Miao:2014nxa,Mozaffar:2016hmg,Lewkowycz:2013nqa} for reviews and references.

It is natural to investigate the minimal surface in more unusual gravity models, since it may provide some important information about the dual field theory. There is an important model, known as C-metric, which is a representative of a class of space-time with boost-rotation symmetry \cite{Weyl:1917gp,Kinnersley:1970zw,Dias:2003xp,Dias:2002mi,Podolsky:2002nk,
Hong:2003gx,Podolsky:2003gm,Griffiths:2005qp,Griffiths:2006tk,
Appels:2016uha,Appels:2017xoe}. It was originally found by Weyl in 1917 and rediscovered many times \cite{Weyl:1917gp}. In 1970, Kinnersley and Walker showed that it describes a pair of black holes which accelerate away from each other due to the existence of conical singularities caused by cosmic strings or struts \cite{Kinnersley:1970zw}. C-metric also admits a non-vanishing value of a cosmological constant $\Lambda$. These solutions can be used to investigate the accelerating black holes in dS and AdS backgrounds \cite{Dias:2003xp,Dias:2002mi,Podolsky:2002nk,Hong:2003gx,Podolsky:2003gm,Griffiths:2005qp}. The global causal structure and physical interpretation of parameters in the metric are investigated in \cite{Griffiths:2006tk}. Recently, the authors of \cite{Appels:2016uha,Appels:2017xoe} addressed the problem of describing the thermodynamics of a slowing charged accelerating black hole in AdS space-time. C-metric also has been used to construct localized black holes on UV branes and plasma ball solutions on IR branes \cite{Emparan:1999wa,Emparan:1999fd,Emparan:2009dj}.

However, the field theory interpretation of C-metric are still, to a large extent, open questions. In the present work we plan to shed some light on these questions by studying the minimal surfaces which are homologous to the given boundary regions. It is worth emphasizing that it is still unclear about whether the minimal surface can be interpreted as the HEE, since the accelerating black holes change the asymptotic structure of the space-time and drag local inertial frames. In this paper we only focus on the geometry features of the minimal surface, leaving the detailed field theory analysis to the future research.

The most general form of C-metric is specified by various parameters, including mass, electric/magnetic charge, angular momentum, acceleration, cosmological constant and NUT charge \cite{Griffiths:2005qp}. It contains a rich variety of solutions. In this paper we concentrate on the simplest form, the static uncharged AdS C-metric in spherical-like coordinates, to avoid the overwhelming complexities. This solution is represented by

\begin{equation}
\begin{aligned}
ds^2&=\frac{1}{\Omega^2}\Bigl[ -f(r)dt^2+\frac{dr^2}{f(r)}
+r^2\Bigl(\frac{d\theta^2}{g(\theta)}
+g(\theta)\sin^2\!\theta {d\phi^2}\Bigr)\Bigr],
\end{aligned}
\label{metric}
\end{equation}
where
\begin{equation}
\begin{aligned}
f(r)&=(1-\alpha^2r^2)\Bigl(1-\frac{2 m}{r}\Bigr)+\frac{r^2}{\ell^2}\;,\\
g(\theta)&=1-2m\alpha \cos\theta.
\end{aligned}
\label{fr}
\end{equation}
The conformal factor
\begin{equation}
\Omega=1-\alpha r \cos\theta\;
\end{equation}
will define the conformal infinity/AdS boundary of the space-time.

There are three free parameters in the metric, $m$, $\alpha$, and $\ell$. In the limit $\alpha \rightarrow 0$, the metric \eqref{metric} reduces precisely to the well-known Schwarzschild AdS black hole in spherical coordinates. The $m$ is related to the black hole mass which is identical to the Schwarzschild mass in $\alpha\rightarrow 0$, $\alpha$ measures the magnitude of the acceleration, and $\ell$ is the AdS radius.

In order to maintain the space-time signature, it is necessary that $g(\theta)>0$, meaning $m\alpha<\frac{1}{2}$. The coordinate $\phi$ is taken to be periodic with the range $(0,\frac{2\pi}{K})$. The whole structure of the metric is controlled by the value of $\alpha$. For a small value of acceleration, $\alpha<\frac{1}{\ell}$, the metric describes a \emph{single} accelerating black hole suspended in AdS space-time. For a large $\alpha>\frac{1}{\ell}$, this represents a pair of accelerating black holes separated by a acceleration horizon. The limit case $\alpha=\frac{1}{\ell}$ describes a black hole bounded to a 2-brane in 4-dimensions, which plays an important role in Randall-Sundram model. In the present work we will only consider the situation of $\alpha<\frac{1}{\ell}$.

The conical singularity is discovered by considering a small circle around the axis. Since $\phi$ takes the range $(0,\frac{2\pi}{K})$, near $\theta=0$ we have
\begin{align}
\frac{circumference}{radius}=\lim_{\theta\rightarrow0}\frac{2\pi\sin{\theta}}{K \theta}g(\theta)=\frac{2\pi}{K}(1-2m\alpha).
\end{align}
It is not $2\pi$ and implies the existence of conical singularity. Similarly, near $\theta=\pi$ we obtain
\begin{align}
\frac{circumference}{radius}=\lim_{\theta\rightarrow \pi}\frac{2\pi\sin{\theta}}{K (\pi-\theta)}g(\theta)=\frac{2\pi}{K}(1+2m\alpha).
\end{align}

Unless $am=0$, the two conical singularities can not be removed simultaneously. However, we can still choose an appropriate constant $K$ to eliminate one of them. Denoting $\theta_+=\pi$ and $\theta_{-}=0$, we can regularize the metric at one pole by setting
\begin{align}
K_{\pm}=g(\theta_{\pm})=1\pm 2m\alpha,
\end{align}
leaving the singularity at the other pole.

At $r=0$ the space-time has unbounded curvature, which predicts the existence of the physical singularity hidden behind the black hole horizon $r_h$ determined by $f(r_h)=0$. The black hole thermodynamics of C-metric, including the definition of thermodynamical quantities, first law, Smarr formula, and phase structure, has been studied in \cite{Appels:2016uha,Appels:2017xoe} and will not be reviewed here. From now on we will concentrate on the the minimal surfaces.

Since we are studying the system in spherical coordinates, the subregion $A$ on the boundary can be selected to be bounded by the line of latitude $\theta=\theta_0$. Because of the symmetry, the minimal surface $\Sigma$ which is homologous to $A$ is parameterized as $r(\theta)$. The induced
metric on $\Sigma$ can be written as
\begin{equation}
\begin{aligned}
d\tilde{s}^2=\frac{1}{\Omega^2}\Bigl[\frac{1}{f(r)}\dot{r}^2+r^2\frac{1}{g(\theta)}\Bigr] d\theta^2+\frac{1}{\Omega^2}r^2g(\theta)\sin^2\theta d\phi^2.
\label{metric2}
\end{aligned}
\end{equation}

The dot indicates the derivative with respect to $\theta$. The area of the minimal surface can be represented as
\begin{align}
Area(\Sigma)=\frac{1}{4}\int^{\frac{2\pi}{K_{\pm}}}_0 d\phi \int^{\theta_0}_0  d\theta \frac{r(\theta) \sin\theta}{\Omega^2}\Bigr[\frac{g(\theta)}{f(r)}\dot{r}(\theta)+r(\theta)^2\Bigr]^{\frac{1}{2}}.
\label{HEE}
\end{align}

Treating the above integrand as the Lagrangian of the system, we can solve the form of $r(\theta)$. Although in spherical coordinates we do not have a conserved quantities because of the existence of $\sin \theta$, the equation of motion
can still be derived straightforwardly by choosing the right boundary condition.

Expanding the solution $r(\theta)$ near $\theta=0$, we have
\begin{align}
r(\varepsilon)=r_*+r_2 \varepsilon^2+O(\varepsilon^n),
\end{align}
where
\begin{align}
r_2=\frac{r_*^3+\ell^2(2m-r_*)(\alpha^2r_*^2-1)}{2\ell^2(2m\alpha-1)(\alpha r_*-1)}.
\end{align}

Notice that no matter whether there is a conical singularity along the axis $\theta=0$, $\dot{r}(0)=0$, meaning the minimal surface always remains smooth in the coordinates. In real curved space-time the coordinates of C-metric are orthogonal, so there is no sharp vertex.

In the limit $\alpha\rightarrow 0$, the C-metric reduces to the Schwarzschild AdS black hole. For a boundary region bounded by $\theta=\theta_0$, the minimal surface satisfies the boundary condition $r(\theta_0)=\infty$, since the Schwarzschild AdS boundary is located at $r=\infty$. In the case of $\alpha\neq 0$, the AdS boundary is determined by $\Omega= 0$, so our boundary condition should be $r(\theta_0)=\frac{1}{\alpha\cos {\theta_0}}$, which is controlled by both $\alpha$ and $\theta_0$. The area of the minimal surface is still divergent. Once we fix the value of $\alpha$, for each $\theta_0$ we can choose a $\theta_c$, which is very close to the $\theta_0$, to be the cutoff.

It is reasonable to consider a Schwarzschild AdS black hole first, then we turn on the acceleration to investigate the variation of the minimal surface. However, in C-metric the location of the AdS boundary depends on the value of $\alpha$, so it is hard to find a universal cutoff if the boundary is varying. In the present work we will fix the acceleration $\alpha$ and AdS radius, satisfying $\alpha<\frac{1}{\ell}$, then study the minimal surfaces by varying black hole mass $m$. Without loss of generality, we set $\alpha=0.5$ and $\ell=1.0$ in this paper.

\begin{figure}
\begin{center}
\includegraphics[width=0.4\textwidth]{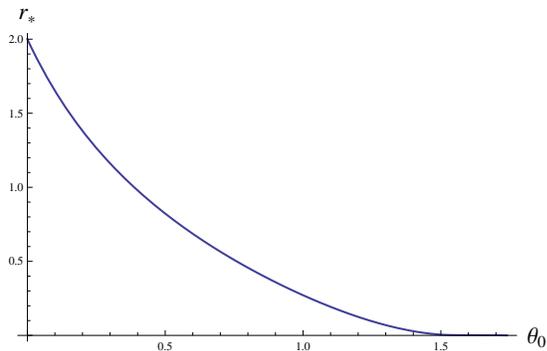}
\caption{Relationship between $r_*$ and $\theta_0$ in the weak field limit $m\rightarrow 0$.}
\label{fig1}
\end{center}
\end{figure}

First we consider the weak field limit $m\rightarrow 0$. The global structure of this case has been well studied in \cite{Podolsky:2002nk,Griffiths:2006tk,Appels:2017xoe}. In Fig.\ref{fig1} we present the relationship between $r_*$ and $\theta_0$. The $r_*$ measures how deep the minimal surface reaches into the bulk. From Fig.\ref{fig1} we obtain that $r_*$ always decreases when $\theta_0$ increases. This brings no surprise. According to the "subregion-subregion" duality, the radial direction, which is regarded as a renormalization scale, can serve as a measure of how well CFT representations in the bulk are protected from local erasures on the boundary \cite{Almheiri:2014lwa,Pastawski:2015qua}. Giving a subregion $A$ on the boundary, we can define an wedge, which is the bulk domain whose boundary is the union of $A$ and the minimal surface $\Sigma$. A bulk field in the wedge can be straightforwardly described on the CFT by the smearing function \cite{Hamilton:2006az,Morrison:2014jha}. It is natural to expect a larger $A$ will access more information about the bulk space-time.

Then we turn on the mass $m$. Notice that the thermal AdS with the same temperature may have the lower Gibbs free energy. However, the existence of the conical singularity prevents the phase transition from accelerating small black hole to the thermal AdS space-time \cite{Appels:2016uha}. As the cutoff $\theta_c$ approaches $\theta_0$, the area of the minimal surface becomes divergent. In the usual way, we can define a regularized area of the minimal surface by subtracting the divergent part corresponding to the weak field limit $m\rightarrow 0$. In the $\alpha\rightarrow 0$ limit, the regularized area measures the entanglement of the excited states relative to the vacuum states \cite{Bhattacharya:2012mi,Sun:2016til,Johnson:2013dka,Caceres:2015vsa}. However, in C-metric the range of the coordinate $\phi$ is also dependent on the $m$. When we vary $m$, the range of $\phi$, the area of region A, and $\partial A$ change simultaneously. We can not find a well defined regularized area of the minimal surface to study the system.

\begin{figure}
\begin{center}
\includegraphics[width=0.4\textwidth]{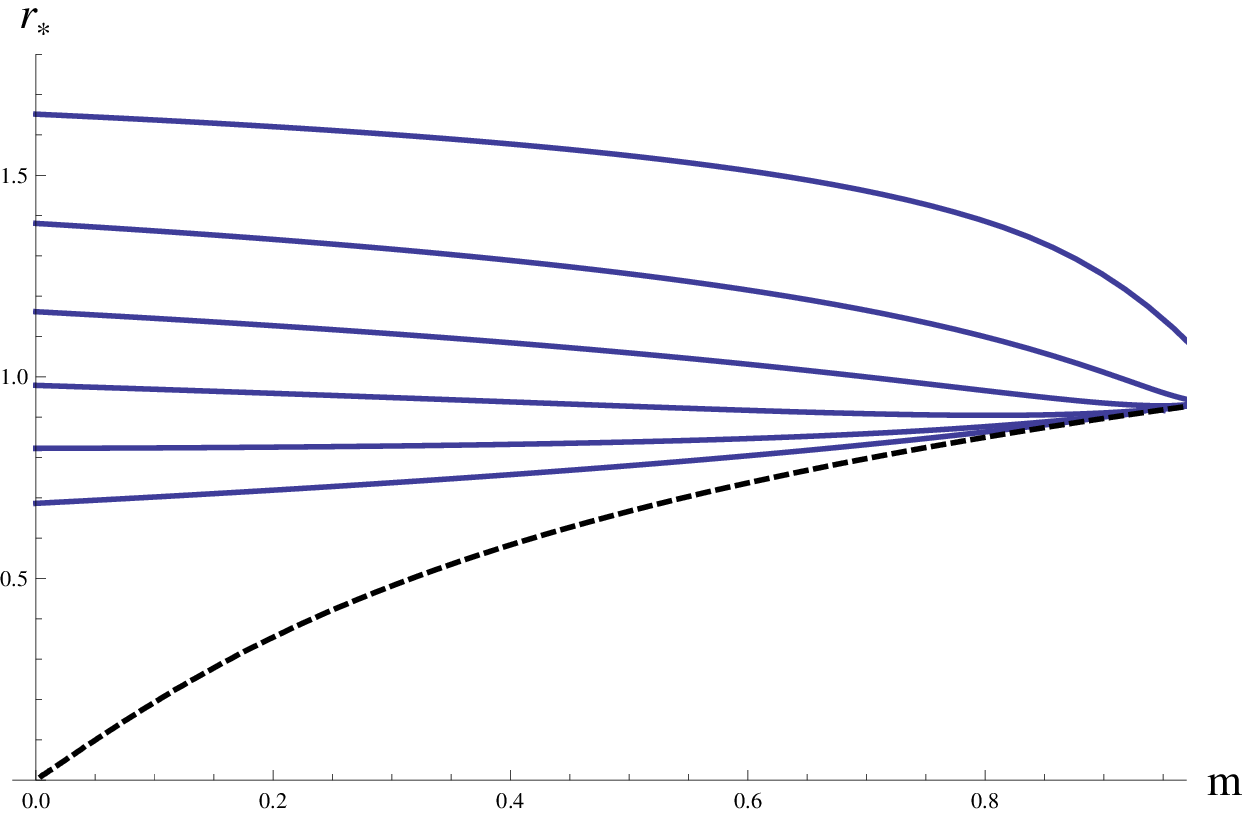}
\includegraphics[width=0.4\textwidth]{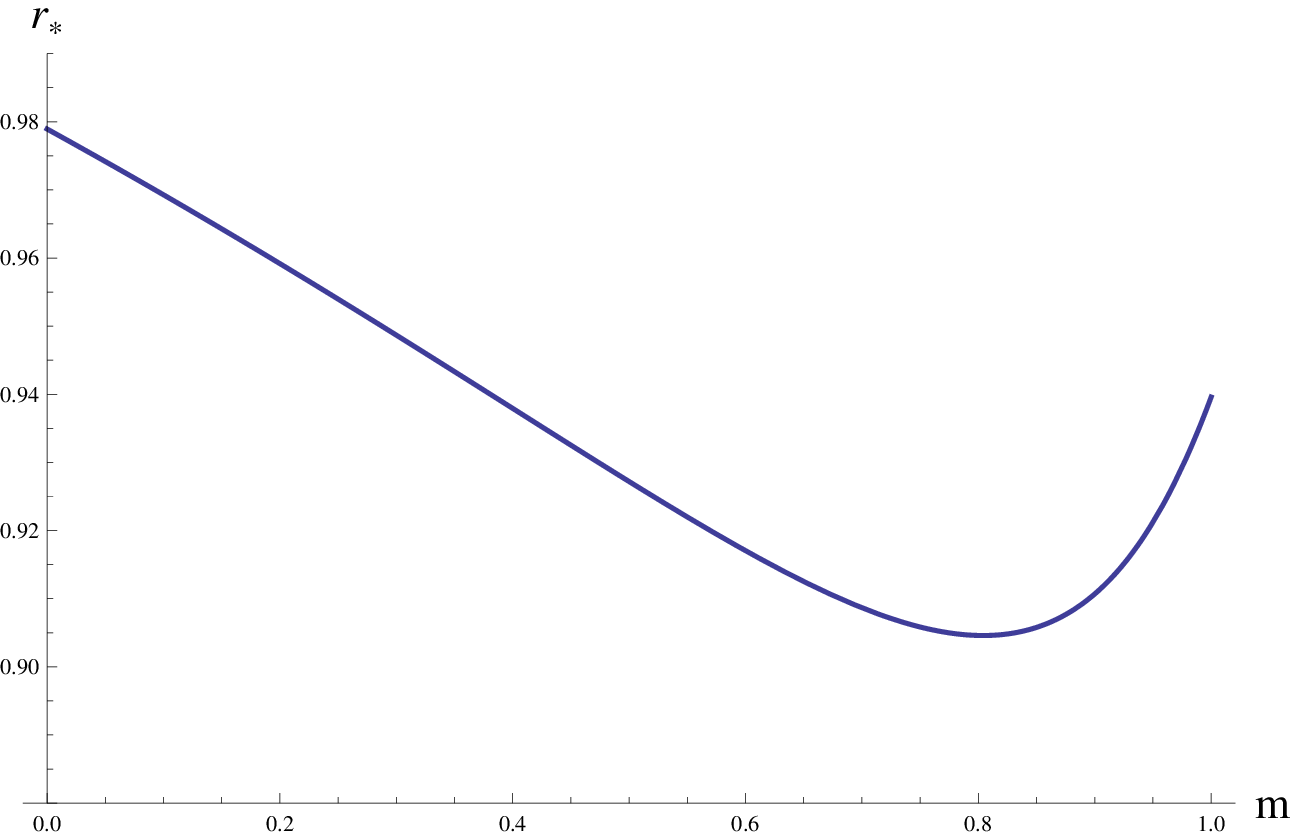}
\caption{Relationship between $r_*$ and $m$ for different values of $\theta_0$.}
\label{fig23}
\end{center}
\end{figure}

However, we can still investigate the $r_*$, since the range of $\phi$ does not affect the equation of motion. In Fig.\ref{fig23} we present the relationship between $r_*$ and $m$ at some fixed $\theta_0$. In the first figure from top to bottom the solid lines correspond to $\theta_0=0.1,0.2,0.3,0.4,0.5,0.6$ respectively, and the dashed line is the black hole event horizon. When the value of $\theta_0$ is small, such as $\theta_0=0.1$, the $r_*$ will decrease as $m$ increases, meaning the minimal surface reaches deeper into the bulk. This is unusual. In the normal case adding energy in the bulk causes the wedge of the subregion A to recede towards the boundary, so the wedge can access fewer bulk field as the energy increases. In our case, the wedge can see further in excited states than in the vacuum states if $\theta_0$ is small enough. When $\theta_0$ becomes larger, the $r_*$ can no longer always decrease. Since the minimal surface cannot penetrate the event horizon, $r_*$ must satisfies $r_*>r_h$. The second figure is a magnification of the case $\theta_0=0.4$. We can see $r_*$ admits a minimal value, predicting two black holes may share the same $\theta_0$ and $r_*$. When $\theta_0$ is large enough, the $r_*$ increases with $m$.

\begin{figure}
\begin{center}
\includegraphics[width=0.4\textwidth]{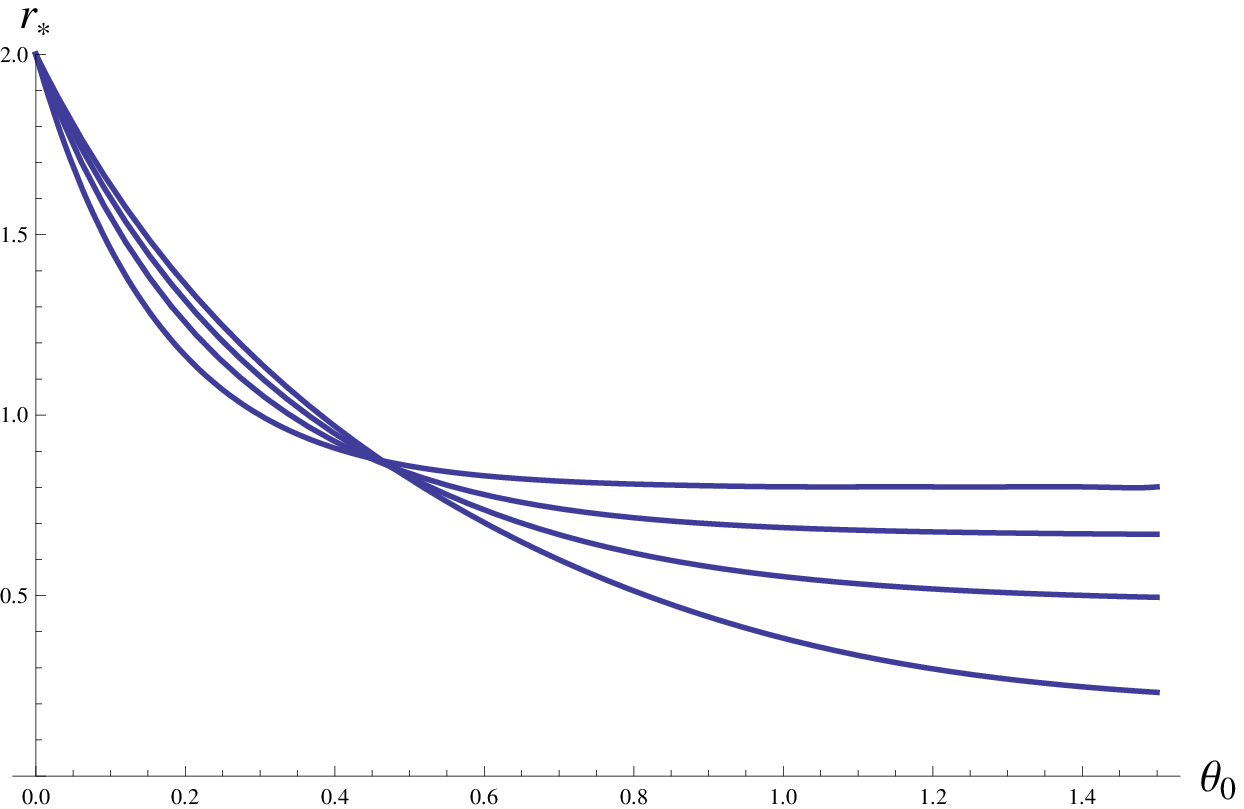}
\includegraphics[width=0.4\textwidth]{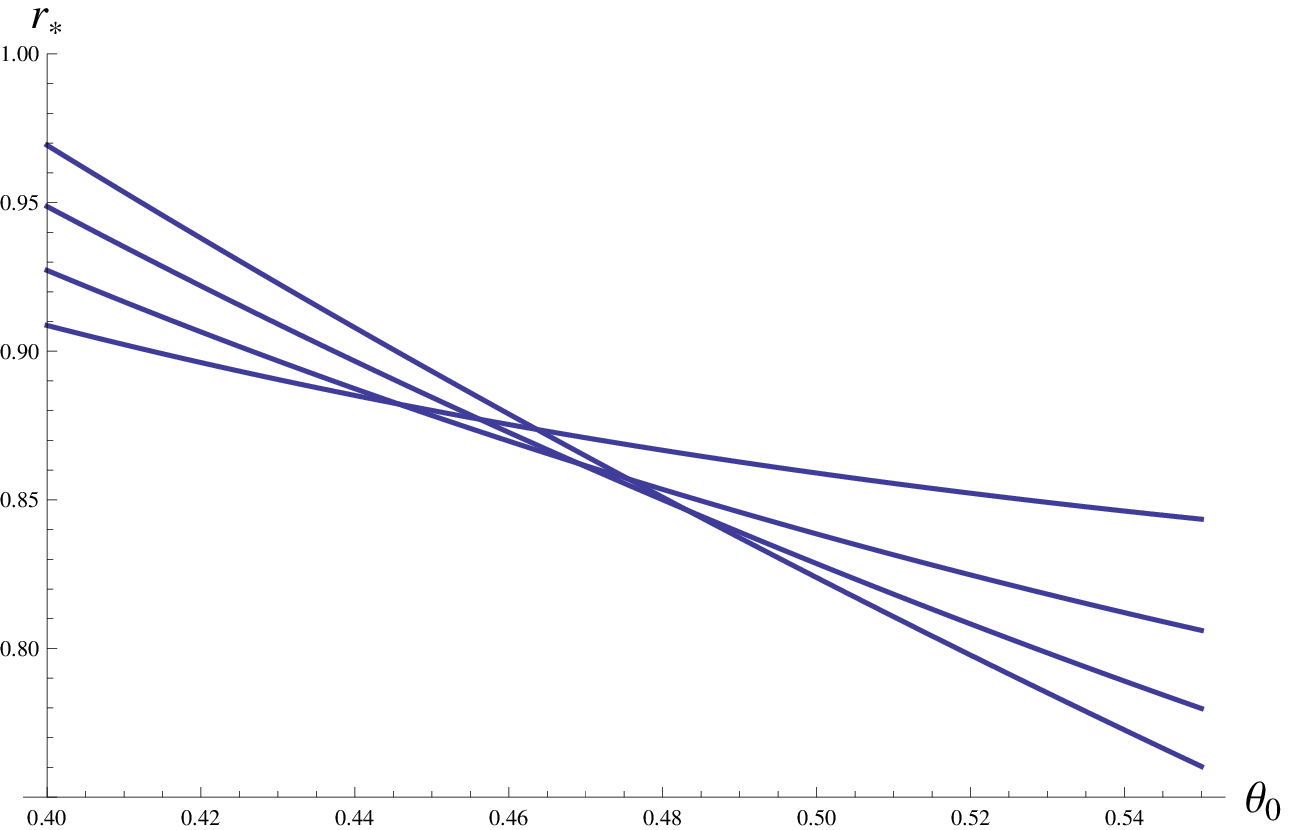}
\caption{Relationship between $r_*$ and $\theta_0$ for different values of $m$.}
\label{fig45}
\end{center}
\end{figure}

In Fig.\ref{fig45} we study the relationship between the $r_*$ and $\theta_0$ at some fixed $m$. For any values of $m$, when $\theta_0\rightarrow 0$, $r_*\rightarrow2$(the AdS boundary). As $\theta_0$ becomes larger, the $r_*$ always decreases, which means larger A accesses more information in the bulk. The four curves in the first figure correspond to $m=0.1,0.3,0.5,0.7$ respectively. For smaller $m$ the curves decrease more slowly at small $\theta_0$ but reach deeper into the bulk at large $\theta_0$. The curves also have intersection points which are close to each other. The second figure is a magnification of the part of intersection points.

Since we cannot find a regularized area of the minimal surface, a new quantity $D(m,\theta_0)$, which is defined to be the minimal surface divided by the area of the subregion A, can be introduced. Similarly definition has also been used in other cases, such as the holographic thermalization process \cite{Balasubramanian:2011ur}. The term $K_{\pm}$, which is contained in the area of A, will be canceled out in $D(m,\theta_0)$. The value of $D(m,\theta_0)$ is divergent and independent of the location of the conical singularity. Now we can define the regularized $D(m,\theta_0)$ by subtracting the part of the weak field limit: $\delta D(m,\theta_0)=D(m,\theta_0)-D(m\rightarrow 0,\theta_0)$. Notice that although the $D(m,\theta_0)$ seems to seems to possess property of some kind of ``density", it is still a non-local quantity. For more details, see \cite{Balasubramanian:2011ur}.

\begin{figure}
\begin{center}
\includegraphics[width=0.4\textwidth]{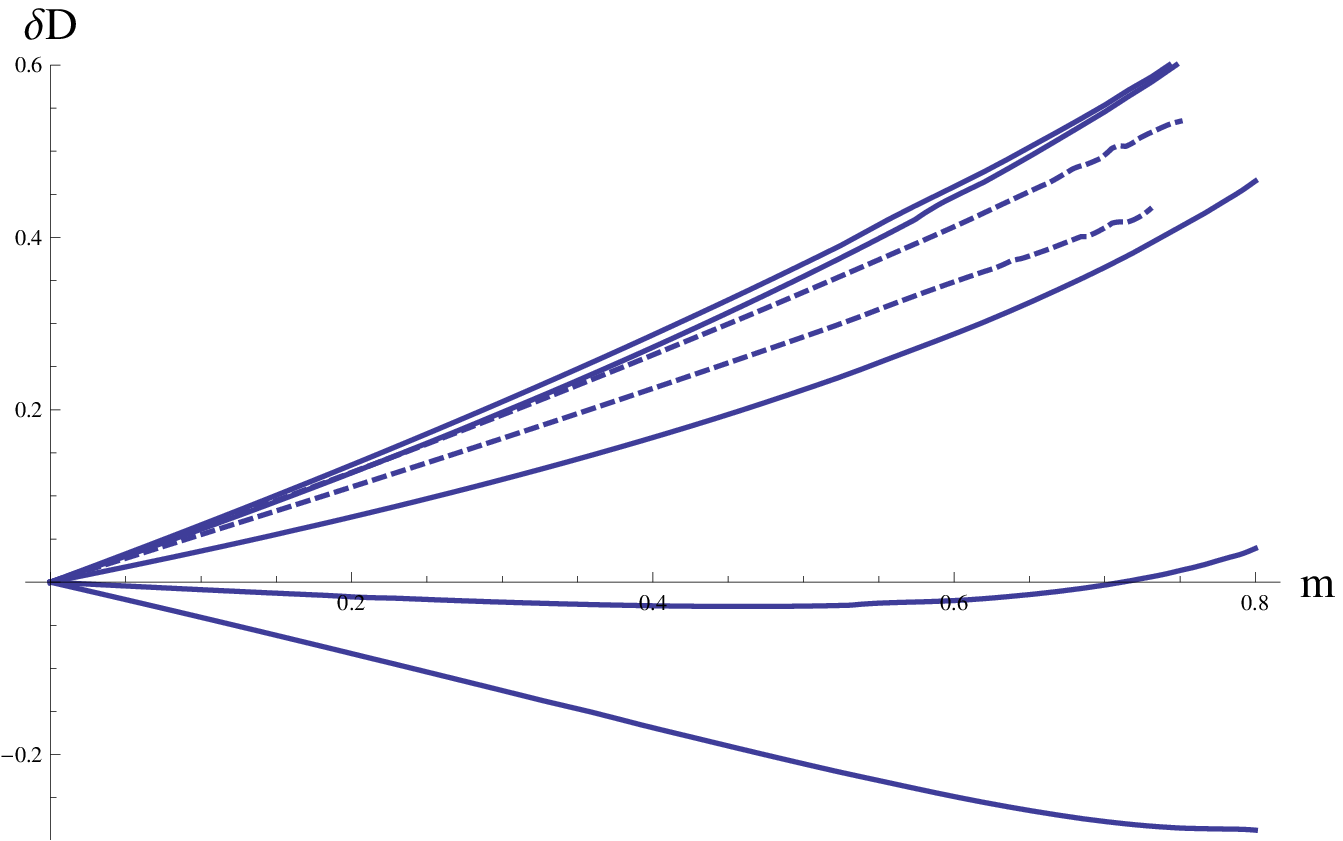}
\includegraphics[width=0.4\textwidth]{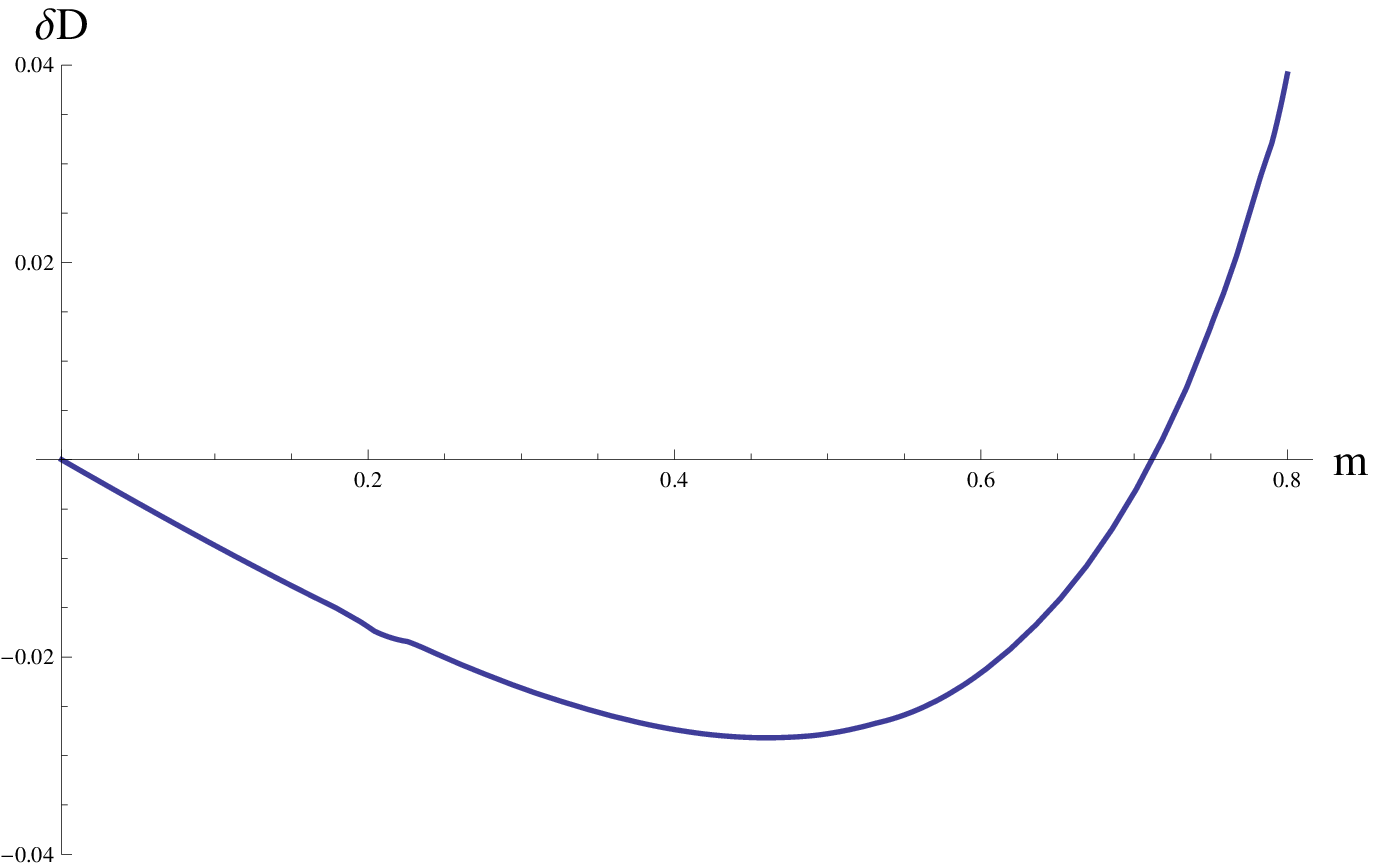}
\caption{Relationship between $\delta D$ and $m$ for different values of $\theta_0$.}
\label{fig67}
\end{center}
\end{figure}

\begin{figure}
\begin{center}
\includegraphics[width=0.4\textwidth]{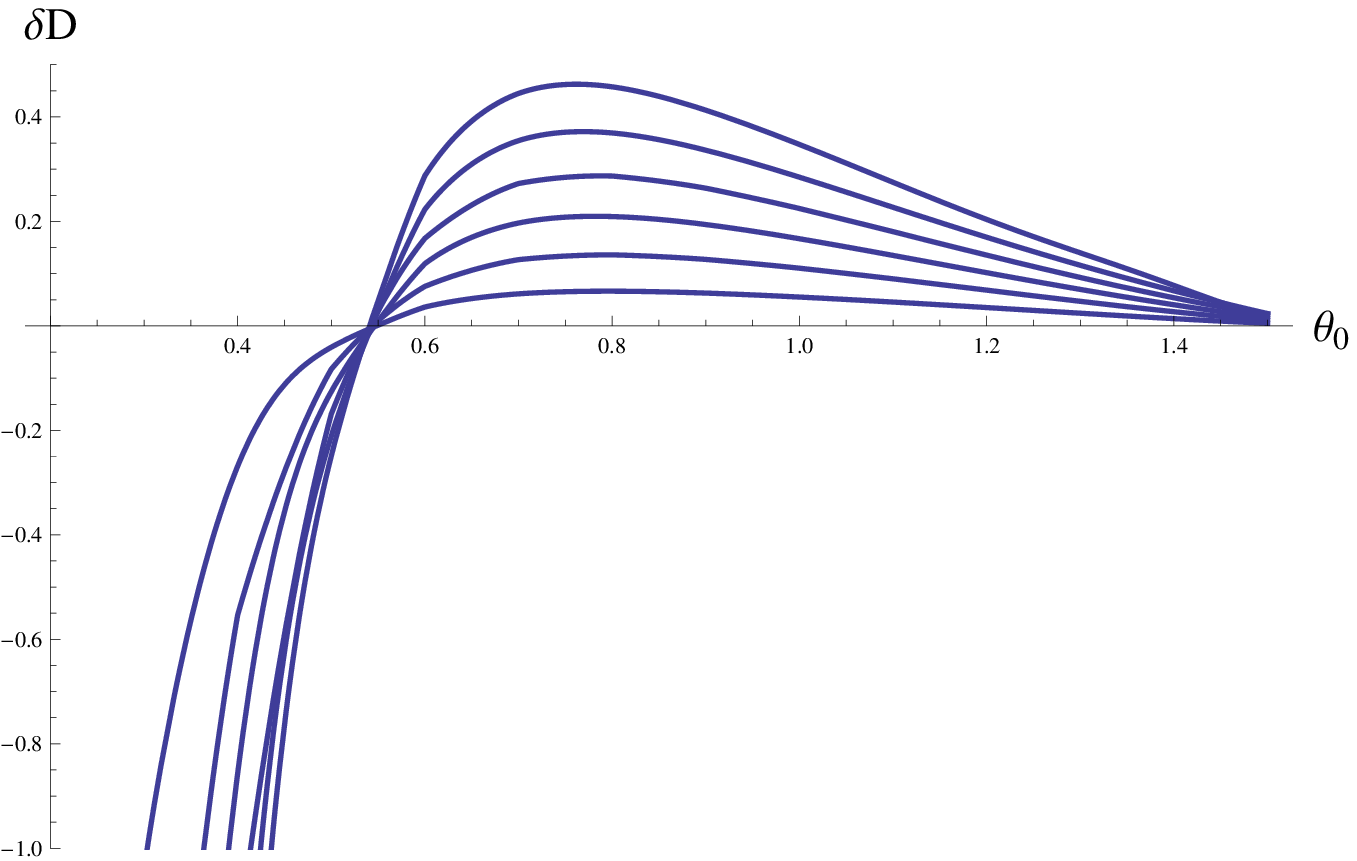}
\includegraphics[width=0.4\textwidth]{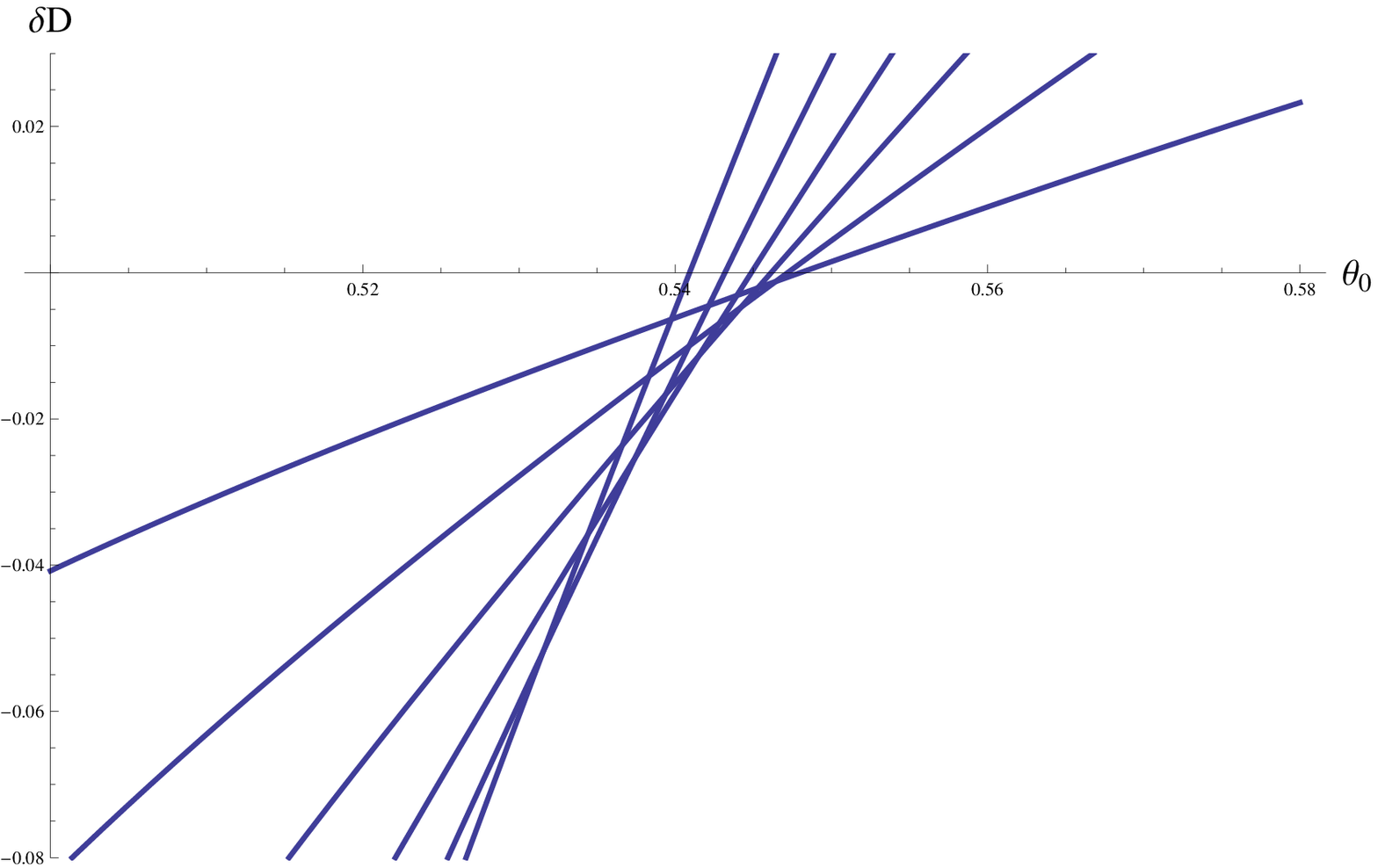}
\caption{Relationship between $\delta D$ and $\theta_0$ for different values of $m$.}
\label{fig89}
\end{center}
\end{figure}

In Fig.\ref{fig67} we present the relationship between the $\delta D(m,\theta_0)$ and $m$ at some fixed $\theta_0$. In the first figure from bottom to top the solid lines correspond to $\theta_0=0.5,0.53,0.6,0.7,0.8$ respectively. We obtain that $\delta D(m,\theta_0)$ decreases in small $\theta_0$. The existence of the conical singularity seems reduce the $\delta D(m,\theta_0)$ near the axis $\theta=0$. When $\theta_0$ becomes larger, the $\delta D(m,\theta_0)$ will increase. In the second figure we present the magnification of the case $\theta_0=0.53$ and find the line admits a minimal value, meaning two black holes may share the same $\delta D(m,\theta_0)$ and $\theta_0$. In the first figure from top to bottom the two dashed lines correspond to $\theta_0=0.9$ and 1.0. The $\delta D(m,\theta_0)$ has a maximal value for a fixed $m$.

In Fig.\ref{fig89} we present the relationship between $\delta D(m,\theta_0)$ and $\theta_0$ at some fixed values of $m$. In the first figure, from left to right the curves correspond to $m=0.1,0.2,0.3,0.4,0.5,0.6$ respectively. All the curves admit a maximal value and have intersection points presented explicitly in the second figure.

Although in the present work we consider the C-metric in the simplest form, due to the overwhelming complexity of the equation of motion, our analysis still relies on the numerical method. The results indicate that the location of the conical singularity does not affect the equation of motion, and the $\delta D(m,\theta_0)$ near the axis $\theta=0$ is reduced. C-metric provides an important example in studying the holography with unusual boundaries.

Similar analysis may also be extended to more general form of C-metric, where AdS$_4$ C-metrics may provide examples of black funnels and black droplets which can be used as (2+1) boundary black holes and candidate dual to Hartle-Hawking states and non-equilibrium states in strongly coupled field theory \cite{Hubeny:2009ru,Hubeny:2009kz,Mefford:2016res}. Other parameters, such as charge and rotation, can also be included. When the equation of motion becomes too complicated, calibration approach may serve as a potential alternative method to obtain the minimal surface \cite{Bakhmatov:2017ihw}. Another interesting metric is the Ernst solution that describes two oppositely charged black holes accelerating away from each other \cite{Ernst1976}. The metric is free of conical singularities and the acceleration is provided by external electromagnetic field. Ernst solution only exists in $\Lambda=0$ and it may shed some light on the holography beyond AdS \cite{Sun:2016dch,Jiang:2017ecm}. It would also be interesting to consider the thermalization process from excited states to excited states \cite{Xu:2017wvu}. We hope to be able to address these questions in the near future.

\begin{acknowledgments}

We would like to thank the referee whose suggestions and comments helped us in improving the original manuscript. Hao Xu thanks Yuan Sun and Liu Zhao for useful discussions. This work is supported by the National Natural Science Foundation of China under the grant No. 11575088.

\end{acknowledgments}

\providecommand{\href}[2]{#2}\begingroup\raggedright\endgroup

\end{document}